\documentclass[floatfix,reprint, superscriptaddress, aps, showpacs, prb]{revtex4-1}

\pdfoutput=1

\usepackage{graphicx}
\usepackage{dcolumn}
\usepackage{bm}
\usepackage{amsmath,amsbsy,amsfonts,revsymb}
\usepackage{color}
\usepackage{textcomp}
\usepackage{slashbox}
\usepackage[table]{xcolor}
\usepackage{xcolor}
\usepackage{tikz}
\usepackage[leftcaption]{sidecap}
\usepackage{collcell}

\usepackage[utf8]{inputenc}
\def\wn/{\,cm$^{-1}$}
\def\area/{\,cm$^{-2}$}
\def\cubic/{$_\mathrm{c}$}
\def\DM/{Dzyaloshinskii-Moriya}
\def\lnpo/{LiNiPO$_4$}
\def\lcpo/{LiCoPO$_4$}
\def\TN/{$T_{\textrm{N}}$}

\def\edc/{\ensuremath{\mathbf{E}}}
\def\eac/{\ensuremath{\mathbf{E}^\omega}}
\def\bdc/{\ensuremath{\mathbf{H}}}
\def\bac/{\ensuremath{\mathbf{H}^\omega}}
\def\kvec/{\ensuremath{{\bf k}}}
\def\Svec/{\ensuremath{{\bf S}}}
\def\hac/{\ensuremath{\mathbf{H}^\omega}}

\def\kvec/{\ensuremath{{\bf k}}}
\def\Svec/{\ensuremath{{\bf S}}}
\def\ew/{\ensuremath{\mathit{E}^\omega}}
\def\bw/{\ensuremath{\mathit{H}^\omega}}
\def\wavenumber{\ensuremath{\tilde{\omega}}}

\newcommand{\ewu}[1]{\ensuremath{\mathit{E}^\omega_{#1}}}

\newcommand{\kparal}[1]{\ensuremath{\mathbf{k}\!\parallel\!\mathbf{{#1}}}}
\newcommand{\bwu}[1]{\ensuremath{\mathit{H}^\omega_{#1}}}
\newcommand{\bparal}[1]{\ensuremath{\mathbf{H}\!\parallel\!\mathbf{{#1}}}}
\newcommand{\vect}[1]{\ensuremath{\mathbf{#1}}}

\makeatletter
\renewcommand{\p@subsection}{}
\renewcommand{\p@subsubsection}{}
\makeatother
\begin{document}


\title{Spin excitations of magnetoelectric LiNiPO${_4}$ in multiple magnetic phases}

\author{L. Peedu}
\thanks{These authors contributed equally to this work.}
\affiliation{National Institute of Chemical Physics and Biophysics, Akadeemia tee 23, 12618 Tallinn, Estonia}
\author{V. Kocsis}
\thanks{These authors contributed equally to this work.}
\affiliation{RIKEN Center for Emergent Matter Science (CEMS), Wako, Saitama 351-0198, Japan}

\author{D. Szaller}
\thanks{These authors contributed equally to this work.}
\affiliation{Institute of Solid State Physics, Vienna University of Technology, 1040 Vienna, Austria}

\author{J. Viirok}
\affiliation{National Institute of Chemical Physics and Biophysics, Akadeemia tee 23, 12618 Tallinn, Estonia}

\author{U. Nagel}
\affiliation{National Institute of Chemical Physics and Biophysics, Akadeemia tee 23, 12618 Tallinn, Estonia}

\author{T. R{\~o}{\~o}m}
\affiliation{National Institute of Chemical Physics and Biophysics, Akadeemia tee 23, 12618 Tallinn, Estonia}

\author{D. G. Farkas}
\affiliation{Department of Physics, Budapest University of Technology and Economics, 1111 Budapest, Hungary}

\author{S. Bord\'acs}
\affiliation{Department of Physics, Budapest University of Technology and Economics, 1111 Budapest, Hungary}
\affiliation{Hungarian Academy of Sciences, Premium Postdoctor Program, 1051 Budapest, Hungary}

\author{D. L. Kamenskyi}
\author{U. Zeitler}
\affiliation{High Field Magnet Laboratory (HFML-EMFL), Radboud University, Toernooiveld 7, 6525 ED Nijmegen, The Netherlands}

\author{Y. Tokunaga}
\affiliation{RIKEN Center for Emergent Matter Science (CEMS), Wako, Saitama 351-0198, Japan}
\affiliation{Department of Advanced Materials Science, University of Tokyo, Kashiwa 277-8561, Japan}

\author{Y. Taguchi}
\affiliation{RIKEN Center for Emergent Matter Science (CEMS), Wako, Saitama 351-0198, Japan}

\author{Y. Tokura}
\affiliation{RIKEN Center for Emergent Matter Science (CEMS), Wako, Saitama 351-0198, Japan}
\affiliation{Department of Applied Physics, University of Tokyo, Hongo, Tokyo 113-8656, Japan}

\author{I. K\'ezsm\'arki}
\affiliation{Department of Physics, Budapest University of Technology and Economics, 1111 Budapest, Hungary}
\affiliation{Experimental Physics 5, Center for Electronic Correlations and Magnetism,Institute of Physics, University of Augsburg, 86159 Augsburg, Germany}

\date{\today }

\begin{abstract}

Spin excitations of magnetoelectric \lnpo/ are studied by infrared absorption spectroscopy in the THz spectral range as a function of magnetic field  through various commensurate and incommensurate magnetically ordered phases up to 33\,T. 
Six spin resonances and a strong two-magnon continuum are observed in zero magnetic field. 
Our systematic polarization study reveals that some of the excitations are usual  magnetic-dipole active magnon modes, while others are either electromagnons,  electric-dipole active, or magnetoelectric, both electric- and magnetic-dipole active spin excitations. 
Field-induced shifts  of the modes  for all three orientations of the field along the orthorhombic axes allow us to refine the values of the relevant exchange couplings, single-ion anisotropies, and the \DM/ interaction on the level of a four-sublattice mean-field spin model. 
This model also reproduces the spectral shape of the two-magnon absorption continuum, found to be  electric-dipole active in the experiment. 

\end{abstract}
\maketitle

\section{Introduction}

Potential of magnetoelectric (ME) materials in applications relies on the entanglement of magnetic moments and electric polarization \cite{Kimura2003,Fiebig2005,Spaldin2005,Eerenstein2006,Cheong2007,Fiebig2009,Dong2015,Fiebig2016NRM}.
Such an entanglement leads not only to the static ME effect but also to the  optical ME effect.
One manifestation of the optical ME effect is the non-reciprocal directional dichroism, a difference in the absorption with respect to the reversal of light propagation direction \cite{Rikken1997,Rikken2002,Barron2004Book,Arima2008}.
The spectrum of non-reciprocal directional dichroism and the linear static ME susceptibility are related via a ME sum rule \cite{Szaller2014}. 
According to this sum rule the contribution of simultaneously magnetic- and electric-dipole active spin excitations to the linear ME susceptibility grows as $\omega^{-2}$ with $\omega\rightarrow 0$.
Indeed, strong non-reciprocal directional dichroism has been observed at low frequencies, typically  in the GHz-THz range, at spin excitations   in several ME materials \cite{Kezsmarki2011,Bordacs2012,Takahashi2012,Kezsmarki2014,Kezsmarki2015,Kuzmenko2015,Bordacs2015,Okamura2015,Nii2017,Yu2018,Kocsis2018,Viirok2019,Okamura2019}.
Beside the interest in the non-reciprocal effect, the knowledge of the spin excitation spectrum and selection rules, i.e. whether the excitations are ordinary magnetic-dipole active magnons, electromagnons (electric-dipole active magnons \cite{Pimenov2006}), or ME spin excitations (simultaneously magnetic- and electric-dipole active spin excitations), is crucial in understanding the origin of static ME effect.

It is well established that the static ME effect is present in several olivine-type Li$M$PO$_4$ ($M=\mathrm{Mn}$, Fe, Co, Ni) compounds \cite{Mercier1967,Mercier1967ME,Mercier1968,Mercier1969,Rivera1994,Kornev2000,Toft-Petersen2015,Khrustalyov2017}.
\lnpo/ is particularly interesting due to many magnetic-field-induced phases, some with   incommensurate magnetic order, which is unique in the olivine lithium-ortho-phosphate family \cite{Toft-Petersen2011}.
However, little is known about the spectrum of spin excitations and their selection rules.

THz absorption spectroscopy offers an excellent tool to investigate  spin excitation spectra over a broad magnetic field range. 
As compared to the inelastic neutron scattering (INS), only spin excitations with zero linear momentum are probed, but with a better energy resolution.
In addition to excitation frequencies, THz spectroscopy can determined whether the spin excitations are  magnetic-dipole active magnons, electromagnons,  or  ME spin excitations.
This information is essential for developing a spin model that would describe the ground and the low lying excited states of the material.

We studied the spin excitation spectra of \lnpo/ in magnetic field using THz absorption spectroscopy.
In the previous INS works two magnon branches were observed below 8\,meV \cite{Jensen2009,Li2009Ni,Toft-Petersen2011}.
Here we broaden the spectral range up to 24\,meV, which allows us to observe additional spin excitations  
and to  identify the polarization selection rules for the spin excitations.
Using a mean-field model we describe the field dependence of the magnetization and the magnon energies in commensurate phases, from where we refine the values of exchange couplings, single-ion anisotropies, and the \DM/ interaction.
Beside magnons described by the mean-field model few other spin excitations, including two-magnon excitations, are observed.

\begin{figure}
	\centering
	\vspace{-10pt}\includegraphics[width=0.8\linewidth]{{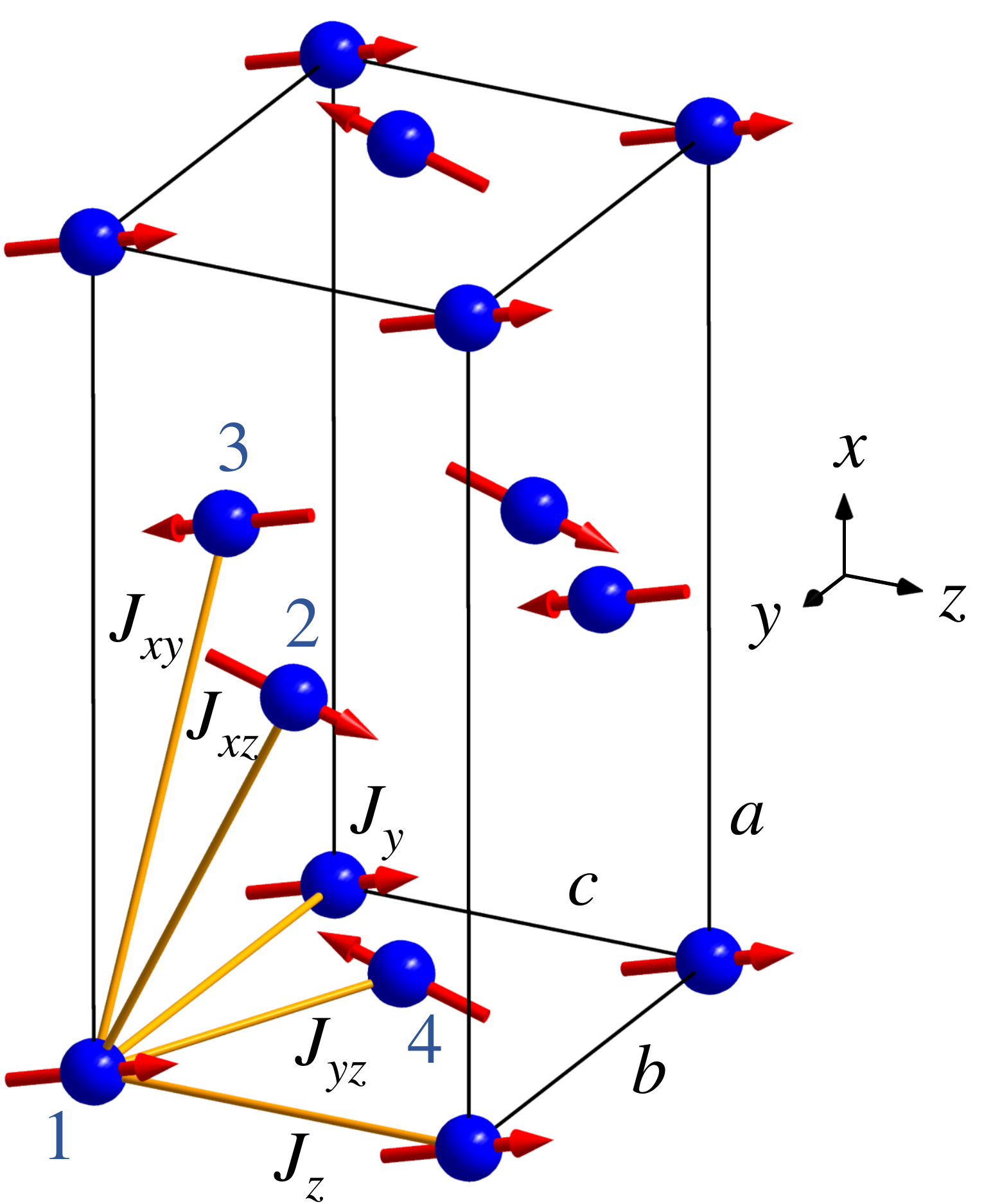}}
	\caption{\label{Fig:Magnetic_cell}
		The ground state spin configuration of \lnpo/  in  zero magnetic field. 
		There are four Ni$^{2+}$  spins, $S=1$,  in the magnetic unit cell drawn as a box. 
		The spins are canted away from the $z$ axis towards the $x$ axis by $\theta=\pm(7.8^\circ \pm 2.6^\circ)$ \cite{Toft-Petersen2011,Jensen2009_12}.
		The numbering of spins and the labeling of exchange interactions corresponds to the spin Hamiltonian described by Eq.\,(\ref{equation:model}).	 
	}
\end{figure}

\begin{figure}
	\centering
	\vspace{-10pt}\includegraphics[width=1\linewidth]{{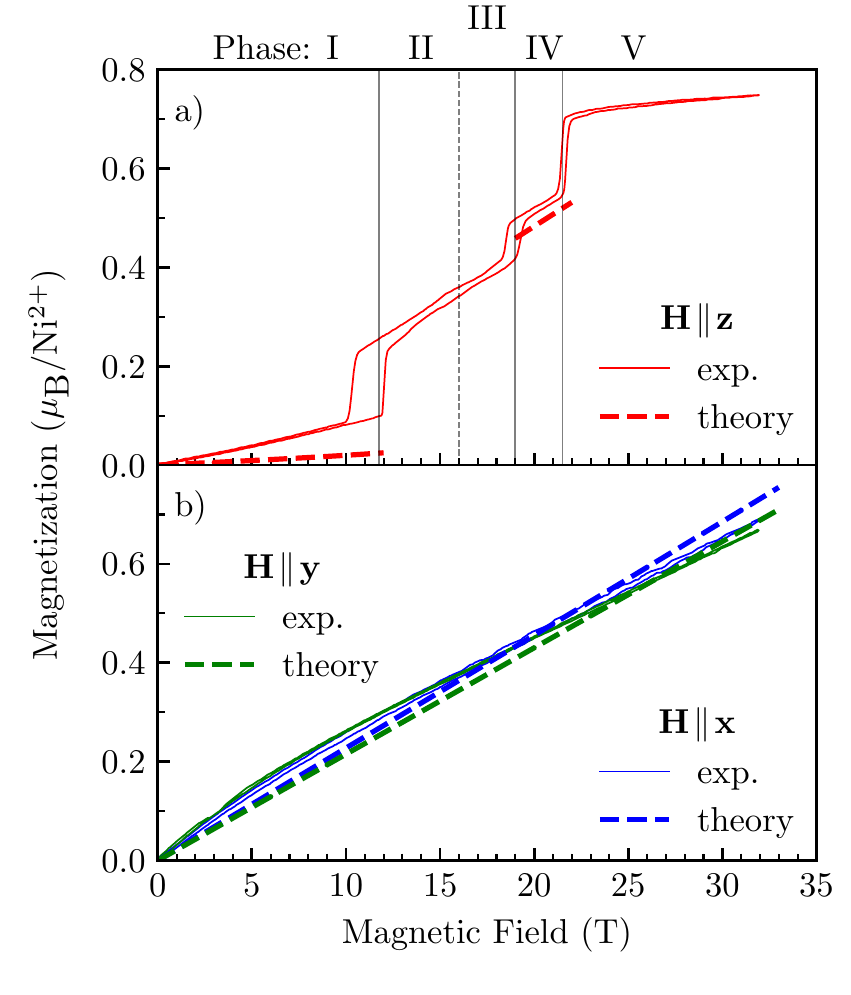}}
	\caption{\label{Fig:Magnetization}
		Magnetic field $\mathbf{H}$ dependence of the magnetization $\mathbf{M}$ parallel to  the field at 2.4\,K. 
		(a)   $\bdc/\parallel \vect{z}$, (b)   $\bdc/\parallel \vect{y}$ (green) and $\bdc/\parallel \vect{z}$ (blue). 
		Solid lines are experimental results and the dashed lines are calculated from  the mean-field model with the parameters of this work listed in Table\,\ref{tabel:exchange parameters}.
	}
\end{figure}

\lnpo/ has orthorhombic symmetry with space group $Pnma$. 
The magnetic Ni$^{2+}$ ion with spin $S=1$ is  inside a distorted O$_{6}$ octahedron.
There are four Ni$^{2+}$  ions in the structural unit cell forming buckled planes perpendicular to the crystal $x$ axis, as shown in Fig.~\ref{Fig:Magnetic_cell}.
The nearest-neighbor spins in the $yz$ plane are coupled by strong AF  exchange interaction which results in a commensurate  AF order below \TN/=20.8$\,$K \cite{Kharchenko2003,Vaknin2004}. 
The ordered magnetic moments are almost parallel to the crystallographic $z$ axis with slight canting towards the $x$ direction \cite{Khrustalyov2004}.
On heating above \TN/ the material undergoes a first-order phase transition into a long-range incommensurate  magnetic structure.  
Further heating results in a second-order phase transition into the paramagnetic state at $T_{\mathrm{IC}} =21.7$\,K, while short-range magnetic correlations persist up to 40\,K \cite{Vaknin2004}.
Owing to the competing magnetic interactions \lnpo/ has a very rich  $H$--$T$ phase diagram with transitions appearing as multiple steps in the field dependence of the magnetization \cite{Khrustalyov2016,Toft-Petersen2017}.
The delicate balance of the nearest-neighbor and the frustrated next-nearest-neighbor exchange interactions puts the material  on the verge of commensurate and incommensurate structures, which  alternate in increasing magnetic field applied along the $z$ axis as shown in Fig.\,\ref{Fig:Magnetization}(a) \cite{Khrustalyov2004,Li2009Ni,Jensen2009}.

\section{Experimental details} 

\lnpo/ single crystals were grown by the floating zone method, similarly as described in Ref.\,[\onlinecite{Baker2011}].
Three samples each with a large face normal to one of the principal axes were cut from the same ingot.
For optical measurements the slabs with thicknesses from 0.87\,mm to 1.09\,mm   had approximately  two degree wedge  to suppress interference caused by internal  reflections. 
Samples were mounted on metal discs where the hole depending on the sample size limited the THz beam cross-section to 8\,--\,16\,mm$^{2}$.

THz measurements up to 17\,T were performed in Tallinn with a Martin-Puplett interferometer and a 0.3\,K silicon bolometer.
High-field spectra from 17\,T up to 33\,T were measured in Nijmegen High Field Magnet Laboratory using a Bruker IFS\,113v spectrometer and a 1.6\,K silicon bolometer.
The experiments above 17\,T were done in the Faraday configuration  ($\mathbf{k}\!\parallel\!\bdc/$), while below 17\,T both the Faraday and the Voigt ($\mathbf{k}\!\perp\!\bdc/$) configuration experiments were performed.
All spectra were measured with an apodized spectral resolution of 0.5\,\wn/.
A linear polarizer was mounted in front of the sample to control the polarization state of the incoming light. 

Absorption was determined by using a reference spectrum. 
The reference spectrum was obtained on the sample in zero magnetic field in the paramagnetic state at $T=30$\,K or by measuring a reference hole with the area equal to  the sample hole area.
In the former case the relative absorption  is calculated as
$
\alpha(H,T) - \alpha(0\text{T},30\text{K}) =  -d^{-1} \ln\left[I(H,\textit{T})/I(0\text{T},30\text{K}) \right]
$
where $d$ is the sample thickness and $I$ is the measured intensity.
In the latter case the absolute absorption is calculated as
$
\alpha = -d^{-1} \ln (I/I_r)$
where  $I_{r}$ is the intensity through the reference hole.

Magnetization up to 32\,T  was measured in Nijmegen High Field Magnet Laboratory on a  Bitter magnet with a vibrating-sample magnetometer (VSM) and additional low field measurements were done using a 14\,T PPMS with VSM option (Quantum Design).

\section{Mean-field model and magnons}\label{sec:Mean-field model}

The terms included in the spin Hamiltonian, exchange interactions, single-ion anisotropy terms, and the Zeeman energy, correspond to those also considered in earlier works on \lnpo/ \cite{Li2009Ni,Jensen2009,Toft-Petersen2011}.
The model contains four spin variables as classical vectors $\{\mathbf{S}_1,\mathbf{S}_2,\mathbf{S}_3,\mathbf{S}_4\}$ in accordance with the four crystallographically non-equivalent positions of the spin $S\,=\,1$  Ni$^{2+}$ ions in \lnpo/. 
The four spins of the magnetic unit cell are connected by five different exchange couplings as presented in Fig.\,\ref{Fig:Magnetic_cell}.
Two of these couplings, $J_y$ and $J_z$, connect spins at the same crystallographic sites producing, irrespective of the spin state, a constant energy shift in the $\Gamma$ point within the four-sublattice model.
Although these terms are omitted in the  analysis of single-magnon excitations, they become relevant in the analysis of two-magnon excitations as discussed in Sec.\,\ref{sec:TwoMagnon}. 
The spin Hamiltonian of the magnetic unit cell in the four-sublattice model reads 

\begin{eqnarray}
	\mathcal{H} &=& \sum_{i=1}^{4}\left[\Lambda_x\left(\vphantom{S_i^y}S_i^x\right)^2 + \Lambda_y\left(S_i^y\right)^2  -g\mu_{\mathrm B}\mu_{0}\mathbf{H}\cdot\mathbf{S}_i \right] \nonumber \\
	&+& 4 \left[ J_{xz}\left(\mathbf{S}_{1}\cdot\mathbf{S}_{2}+\mathbf{S}_{3}\cdot\mathbf{S}_{4}\right)  
	+ J_{xy}\left(\mathbf{S}_{1}\cdot\mathbf{S}_{3}+\mathbf{S}_{2}\cdot\mathbf{S}_{4}\right) \right. \nonumber \\
	&+& \left. J_{yz}\left(\mathbf{S}_{1}\cdot\mathbf{S}_{4}+\mathbf{S}_{2}\cdot\mathbf{S}_{3}\right)  \right.\nonumber \\
	&+& \left.  D_y\left(S_1^z S_4^x - S_1^x S_4^z + S_3^z S_2^x - S_3^x S_2^z  \right)\right].
	\label{equation:model}
\end{eqnarray}

Due to the strongly distorted ligand cage of the magnetic ion, the orthorhombic anisotropy of the crystal is taken into account by two single-ion hard-axis anisotropies, $\Lambda_x, \Lambda_y>0$. 
The parameters in the  Zeeman term  are the $g$ factor $g$, the Bohr magneton $\mu_{\mathrm{B}}$, and the vacuum permeability $\mu_{0}$. 
Parameters $J_{xz}$, $J_{xy}$, and $J_{yz}$ are the isotropic Heisenberg exchange  couplings as shown in  Fig.\,\ref{Fig:Magnetic_cell}, while $D_y$ is the \DM/ interaction. 

According to the neutron diffraction studies \cite{Santoro1966,Vaknin1999} the ground-state spin configuration of \lnpo/ in zero magnetic field is a predominantly collinear AF order,  where $\mathbf{S}_1$ and $\mathbf{S}_2$ point in $+z$, while $\mathbf{S}_3$ and $\mathbf{S}_4$ in $-z$ direction, shown in Fig.\,\ref{Fig:Magnetic_cell}.
Thus, the dominant exchange interaction is the AF $J_{yz}>0$ coupling, while $z$ is an easy axis as $\Lambda_x$, $\Lambda_y>0$. 
On top of the collinear order a small alternating canting of spins with net spin along $x$ is superimposed \cite{Jensen2009_12}.
Canting is  induced by  the \DM/ coupling $D_y$ and breaks the equivalence of $\mathbf{S}_1$ and $\mathbf{S}_2$ as well as $\mathbf{S}_3$ and $\mathbf{S}_4$.
The canting angle  $\theta$ measured from the $z$ axis is approximately
\begin{equation}
\tan\theta \approx \frac{2D_y}{\Lambda_x - 4(J_{xz}-J_{yz})}.
\label{equation:spins_angle}
\end{equation}
At each magnetic field, the ground-state spin configuration is obtained by minimizing the energy corresponding to Eq.\,(\ref{equation:model}). 

The resonance frequencies and amplitudes of modes are calculated using the Landau-Lifshitz equation \cite{white2007}
\begin{equation}\label{eq:LL_S}
\dot{\Svec/}_i = -\frac{1}{\hbar}  \Svec/_i \times \frac{\partial \,\mathcal H}{\partial \,\Svec/_i},
\end{equation}
where $\dot{\Svec/}_i	\equiv {\mathrm d} \Svec/_i/{\mathrm d}t$.

We solve Eq.\,(\ref{eq:LL_S}) for small spin deviations  $\{\delta \Svec/  \}\equiv \{\delta \Svec/_1, \ldots  \delta \Svec/_N\}$ from the equilibrium $\{ \Svec/^0  \}\equiv \{ \Svec/_1^0, \ldots   \Svec/_N^0\}$, where $\{\Svec/\}=\{\Svec/^0\}+\{\delta \Svec/\}  $, with $N$ spins in the magnetic unit cell.
It follows from Eq.\,(\ref{eq:LL_S}) that  $\delta \Svec/_i\perp \Svec/_i$, leaving the spin length constant in the first order of $\delta \Svec/_i$.
Inserting $\{\Svec/\}$ into Landau-Lifshitz equation Eq.\,(\ref{eq:LL_S}) and keeping only terms linear in $\delta \Svec/_i$ (terms zero-order in $\delta \Svec/_i$ add  up to zero) we get
\begin{equation}\label{eq:LL_small_displacements}
\delta\dot{\Svec/}_i = -\frac{1}{\hbar} \Svec/_i^0 \times \frac{\partial \,\mathcal H_\delta}{\partial \,\Svec/_{i}},
\end{equation}
where the effective field is

\begin{equation}
\frac{\partial \,\mathcal H_\delta}{\partial \,\Svec/_{i}} = \frac{\partial \,\mathcal H}{\partial \,\Svec/_{i}}\Big|_{\{\Svec/^0\} + \{\delta \Svec/ \}}.
\end{equation}

We solve Eq.\,(\ref{eq:LL_small_displacements}) by assuming harmonic time dependence $\delta \Svec/_i^{}(t) =\delta \Svec/_i^{}\exp(i\omega t) $.
The number of modes is equal to the number of spins in the unit cell.

To calculate the absorption of electromagnetic waves by the magnons we introduce damping.
The Landau-Lifshitz-Gilbert equation \cite{Gilbert2004} for the $i$-th spin is 
\begin{equation}\label{eq:LLG_S}
\dot{\Svec/}_i=- \frac{1}{\hbar}  \vect{S}_i \times \frac{\partial \,\mathcal H_\delta}{\partial \,\Svec/_i}
+ \frac{\alpha}{\hbar S_i}  \vect{S}_i \times  \vect{S}_i \times \frac{\partial \,\mathcal H_\delta}{\partial \,\Svec/_i},
\end{equation}
where  $\alpha$ is a positive dimensionless damping parameter and  small,  $\alpha\ll 1$.
Using $\vect{A}\times\vect{B} \times \vect{C}= \vect{B}(\vect{A}\cdot \vect{C})-\vect{C} (\vect{A}\cdot \vect{B}) $, and adding a weak harmonic alternating magnetic field, $\hac/(t)=\hac/\exp(i\omega t)$, to the effective field yields the following form of the equation of motion up to terms linear in $\delta \Svec/_i$ and $\hac/$:

\begin{eqnarray}\label{eq:LLG_smallAC}
\dot{ \delta\Svec/_i}=& -&\frac{1}{\hbar}    \vect{S}_i^0 \times   \left[\frac{\partial \,\mathcal H_\delta}{\partial \,\Svec/_i}- \mu_0 \hac/(t) \right]\\
&+&   \frac{\alpha}{\hbar } \frac{\vect{S}_i^0}{S_i^0} \vect{S}_i^0 \cdot \left[\frac{\partial \,\mathcal H_\delta}{\partial \,\Svec/_i}- \mu_0 \hac/(t) \right] \nonumber\\
&-&  \frac{\alpha}{\hbar } S_i^0 \left[\frac{\partial \,\mathcal H_\delta}{\partial \,\Svec/_i}- \mu_0 \hac/(t) \right].\nonumber
\end{eqnarray}

The absorption of electromagnetic waves by the spin system related to magnetic dipole excitations is calculated from the Eq.\,(\ref{eq:LLG_smallAC}) by inserting $\delta \Svec/_i^{}(t) =\delta \Svec/_i^{}\exp(i\omega t) $ and $\hac/(t)=\hac/\exp(i\omega t)$.
The frequency-dependent magnetic susceptibility  tensor $\hat{\chi}(\omega)$ is obtained by summing up all the magnetic moments in the unit cell,  $\vect{M}=\gamma \hbar \sum_{i=1}^N\Svec/_i $ in Eq.\,(\ref{eq:LLG_smallAC}), and making a transformation into form 
\begin{equation}\label{eq:susceptibility}
\gamma \hbar\,\, \left[ \sum_{i=1}^M\delta\Svec/_i^{}(t)\right]= \hat{\chi}(\omega)\mu_0 \hac/(t).
\end{equation}

The absorption coefficient is $\alpha_{i,j} =  2\omega c_0^{-1} \mathrm{Im}\, \mathcal{N}_{i,j}$,
where the complex index of refraction   is $\mathcal{N}_{i,j}=\sqrt{\epsilon_{ii}\mu_{jj}}$ assuming small  polarization rotation and negligible linear magnetoelectric susceptibilities $\chi^{em}_{ij}$, $\chi^{me}_{ji}$.
The magnetic permeability is $\mu_{jj}(\omega)=1+\chi_{jj}(\omega)$ and the background dielectric permittivity  is $\epsilon_{ii}$.
The polarization of incident radiation is defined as $\{E_i^\omega, H_j^\omega\}$ where $i$ and $j$ are $x$, $y$, or $z$.
If $\chi_{jj}(\omega)\ll 1$,
\begin{equation}\label{eq:IndexRefr_magnetic}
\mathcal{N}_{i,j}\approx \sqrt{\epsilon_{ii}}\left[1+\frac{\chi_{jj}(\omega)}{2}\right].
\end{equation}

Thus, for real $\epsilon_{ii}$ the absorption is
\begin{equation}\label{eq:abs_mode_n}
\alpha_{i,j}(\wavenumber)=2\pi \wavenumber \sqrt{\epsilon_{ii}}\, \mathrm{Im}\, \chi_{jj}(\wavenumber),
\end{equation}
where units of wavenumber are used, $[\wavenumber_n]=$\wn/. 

The values of magnetic interactions and anisotropies obtained in this work, see Table\,\ref{tabel:exchange parameters}, reproduce the magnetic field dependence of the magnetization, canting angle $\theta$, and frequencies of four spin excitations in the commensurate magnetic phase of \lnpo/.

\def\arraystretch{1.5}

\begin{table}
	
	\begin{ruledtabular}	
		\caption{The parameters of the mean-field model used to describe the static magnetic properties and single- and two-magnon excitations in \lnpo/:
			exchange couplings $J_{ij}$ and $J_{k}$, single-ion anisotropy constants $\Lambda_{i}$, \DM/  coupling $D_y$, and $g$ factor $g$. 
			Units are in meV except the dimensionless $g$. 
		} \label{tabel:exchange parameters}	
		\vspace{5pt} 
		\begin{tabular}{ *{11}{c} }
			$J_{y}$&$J_{z}$&$J_{xz}$ & $J_{xy}$ & $J_{yz}$ & $\Lambda_{x}$ & $\Lambda_{y}$ & $D_y$ & $g$ & Ref. \\ \hline
			0.65&0.16&-0.17 & 0.16 & 1.24 & 0.14 & 0.74 & 0.41 & 2.2 & [$\ast$]  \\ \hline
			0.67&-0.06&-0.11 & 0.32 & 1 & 0.41 & 1.42 & 0.32 & 2.2 & [\onlinecite{Toft-Petersen2011}]  \\ \hline
			0.67&-0.05&-0.11 & 0.3 & 1.04 & 0.34 & 1.82 &  &  & [\onlinecite{Jensen2009}]  \\ \hline
			0.59&-0.11&-0.16 & 0.26 & 0.94 & 0.34 & 1.92 & &  & [\onlinecite{Li2009Ni}]  \\ 
			
		\end{tabular}
	\end{ruledtabular}
\flushleft $\ast$ this work
\end{table}

\section{Experimental results}

The \lnpo/ samples were characterized by measuring the magnetization along the $x$, $y$, and $z$ directions, shown in Fig.\,\ref{Fig:Magnetization}.
The magnetization increases continuously for \bparal{x} and \bparal{y}, while for \bparal{z} there is step at 12\,T, 19\,T, and 21.5\,T.
These steps correspond to magnetic field induced changes in the ground state spin structure. 
The phases I and IV are commensurate, while II, III, and V are incommensurate \cite{Toft-Petersen2011,Toft-Petersen2017}.
The boundary between phases II and III    at 16\,T,   where the periodicity of the incommensurate spin structure changes \cite{Toft-Petersen2011}, is hardly visible in the magnetization data \cite{Khrustalyov2004,Toft-Petersen2017}.
The size of the magnetic  unit cells is the same in phases I and IV \cite{Toft-Petersen2017}, i.e.,   four spins as shown in Fig.\,\ref{Fig:Magnetic_cell}. 

\begin{figure}
	\includegraphics[width=0.5\textwidth]{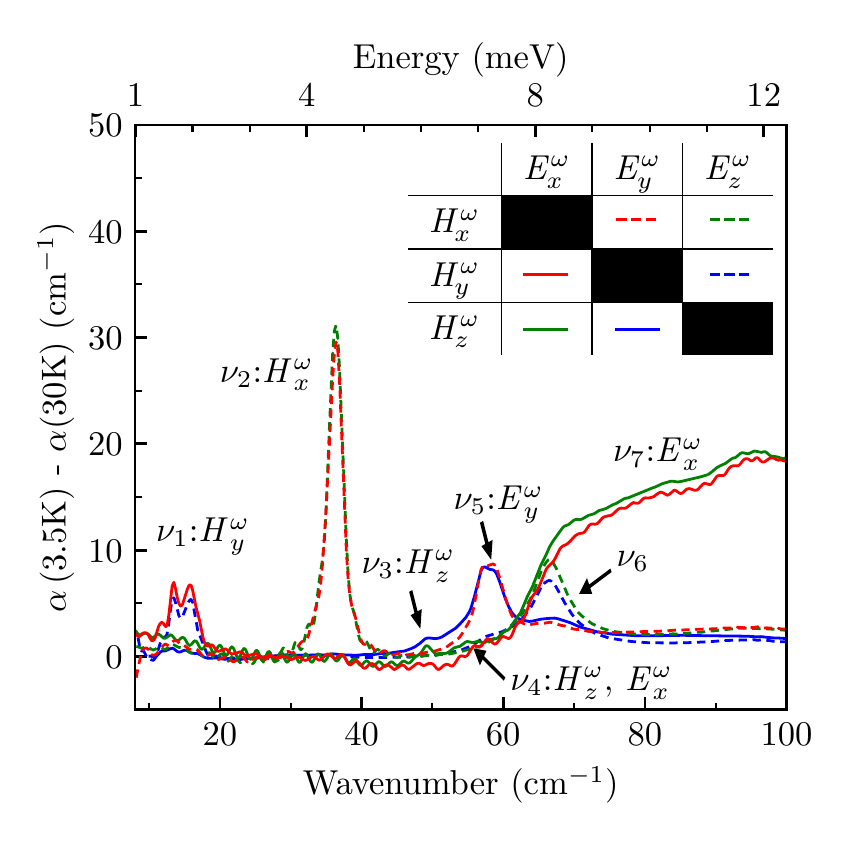}
	\caption{\label{Zero_field} THz absorption spectra of spin  excitations in \lnpo/ in $H=0$ at $T=3.5$\,K.
	Directions of THz radiation propagation are \kparal{x} (blue), \kparal{y} (green) and \kparal{z} (red). 
	Two orthogonal THz radiation polarizations for a given $\mathbf{k}$ vector direction are indicated by solid and dashed lines. 
	Directions of the oscillating THz fields $\{E^\omega_i, H^\omega_j\}$ are indicated in the inset.
	$\nu_n$ labels the modes, $n=1,\ldots, 7$, with $H^\omega_j$ or $E^\omega_i$ indicating the magnetic- or electric-dipole activity of the mode, respectively.
	$\nu_4$ and  $\nu_6$ are ME excitations (for the characterization of $\nu_6$   see Table\,\ref{Table:nu6_excite_conf}).}
\end{figure}

\begin{SCtable}[50]
	\setlength{\tabcolsep}{0pt}
	\centering
	\caption{The  excitation configurations of ME mode $\nu_{6}$. 
		The area of the symbol is approximately proportional to the absorption line area. 
		The color coding is the same as in Fig.\,\ref{Zero_field}.}
	\label{Table:nu6_excite_conf}
	{
		\newcolumntype{C}[1]{@{}>{\centering\arraybackslash\hspace{0pt}}p{#1}@{}}
		\begin{tabular}{C{2em}|C{2em}|C{2em}|C{2em}}
			$\nu_6$& \ewu{x} & \ewu{y} & \ewu{z} \\ 
			\hline 
			\bwu{x} & \cellcolor{black!25} & \tikz\draw[red,fill=white,style=densely dashed] (0,0) circle (0.5ex); &  \tikz\draw[black!60!green,fill=white,style=densely dashed] (0,0) circle (1ex); \\ 
			\hline 
			\bwu{y} &  &\cellcolor{black!25}& \tikz\draw[blue,fill=white,style=densely dashed] (0,0) circle (1ex); \\ 
			\hline 
			\bwu{z} & \tikz\draw[black!60!green,fill=white] (0,0) circle (0.5ex); & \tikz\draw[blue,fill=white] (0,0) circle (0.5ex); & \cellcolor{black!25} \\ 
	\end{tabular} }
\end{SCtable}

\makeatletter\onecolumngrid@push\makeatother
\begin{figure*}[!thbp]
	\includegraphics[width=1\textwidth]{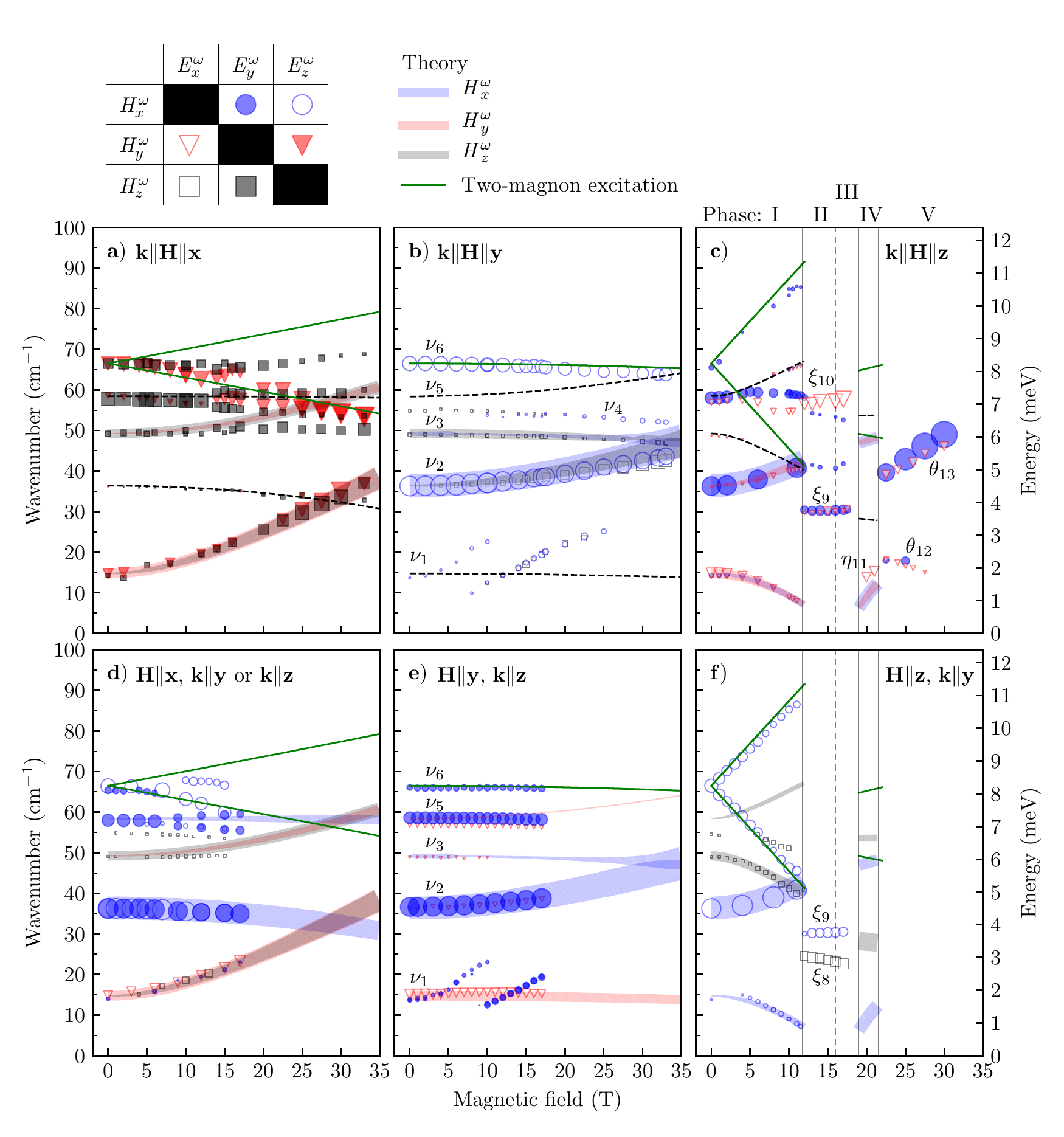}
	\caption{\label{Fig:Magnetic_spectra}
		Magnetic field dependence of the magnon resonance frequencies and absorption line areas at $T=3.5\,\mathrm{K}$. 
		Panels (a), (b), and (c) correspond to measurements in the Faraday  ($\kvec/\parallel \bdc/$), while panels (d), (e), and (f) correspond to experiments in the Voigt  ($\kvec/\perp \bdc/$) configuration.
		The direction of magnetic field is (a), (d) -- \bparal{x}, (b), (e) -- \bparal{y}, and (c), (f) -- \bparal{z}.  
		Symbols are the fit results of experimentally measured resonances and correspond to six combinations of  $\{E^\omega_i, H^\omega_j\}$ as indicated on top of the figure. 
		The symbol area is proportional to the  experimental absorption line area.
		The solid lines are the results of the model calculations based on Eq.\,(\ref{equation:model})--(\ref{eq:LLG_smallAC}).
		The width of the line is proportional to the square root of the   line area calculated  in the magnetic dipole approximation.		
		The color of the symbol and the line is determined by the magnetic component of light, \bwu{x} -- blue, \bwu{y} -- red, and \bwu{z} -- black.
		The line positions of modes with vanishing theoretical intensity in all  measured  configurations of panels (a), (b), and (c) are shown by black dashed lines.
		The green solid line is the two-magnon excitation $\nu_6$.
		The phase boundaries determined from the magnetic field dependence of the THz spectra are shown by vertical solid  lines in (c) and (f); the phase boundary between II and III,  vertical dashed line,   is from Ref.\,[\onlinecite{Khrustalyov2004,Toft-Petersen2017}].}
\end{figure*}
\clearpage
\makeatletter\onecolumngrid@pop\makeatother

The zero-field THz absorption spectra measured at 3.5\,K are shown in Fig.\,\ref{Zero_field}.
Three absorption lines are identified as magnetic-dipole active magnons: $\nu_1=16.1$\wn/, $\nu_2=36.2$\wn/, and $\nu_3=48.4$\wn/. 
The excitation $\nu_5=56.4$\wn/ is an $E_y^\omega$-active electromagnon.
The excitations $\nu_4=54.8$\wn/ and $\nu_6=66.4$\wn/ are ME spin excitations;
$\nu_4$ is $\{E^\omega_x,H^\omega_z\}$-active, while  $\nu_6$ is present in five different combinations of oscillating electric and magnetic fields with the strongest intensity in $E^\omega_z$ polarization, see Table\,\ref{Table:nu6_excite_conf}.
There is an $E^\omega_x$-active  broad absorption band $\nu_7$. 

All seven modes $\nu_1, \ldots ,\nu_7$ are absent above \TN/. 
Since no sign of structural changes has been found in the neutron diffraction \cite{Jensen2009_12} and in the spectra of Raman-active phonons at \TN/ \cite{Fomin2002}, the lattice vibrations can be excluded and all new  modes are assigned to spin excitations of \lnpo/. 

The magnetic field dependence of resonance frequencies and absorption line areas is presented in Fig.\,\ref{Fig:Magnetic_spectra} as obtained from the fits of the absorption peaks with the Gaussian line shapes. 
When the magnetic field is applied in  \bparal{x} or \bparal{y} directions, Fig.\,\ref{Fig:Magnetic_spectra}(a) or (b), we found a continuous evolution of modes up to the highest field of 33\,T.
On  contrary, for \bparal{z} we observed discontinuities in the spin excitation frequencies, approximately at 12\,T, 19\,T and 21.5\,T.
These fields correspond   to the field values where  the steps are seen in the magnetization in Fig.\,\ref{Fig:Magnetization}. 
The boundary between II and III at 16\,T is not visible in the THz spectra.
Apparently the spin excitation spectra are rather insensitive to the change of the magnetic unit cell size   within the incommensurate phase.

The mean-field model (Sec.\,\ref{sec:Mean-field model}) predicts four magnon modes for a four sub-lattice system and they are assigned to $\nu_{1}$, $\nu_{2}$,  $\nu_{3}$ and $\nu_{5}$.
The magnetic field dependence and the selection rules of the magnetic-dipole active magnons $\nu_{1}$, $\nu_{2}$ and $\nu_{3}$ are reproduced well by the mean-field model, Fig.\,\ref{Fig:Magnetic_spectra}. 
However, only the  energy of the magnon   $\nu_{5}$ is reproduced by the model and not the intensity as this excitation is found to be an electromagnon in the experiment.

The resonances $\nu_{4}$, $\nu_{6}$, and the band $\nu_{7}$ cannot be described within the four-sublattice mean-field model.
The weak $\nu_{4}$ mode is a ME spin excitation, $\{E^\omega_x,H^\omega_z \}$ active, which might be related to a spin-stretching excitation allowed for $S>1/2$ \cite{Penc2012}.
The $\nu_{6}$ mode is a ME two-magnon excitation and $\nu_{7}$ an $E^\omega_x$-active two-magnon excitation band, as will be discussed below.
The excitations $\xi_{8}, \xi_9, \xi_{10}$, and $\theta_{12}, \theta_{13}$ are only present in the incommensurate phases II, III, and in phase V with more than four spins per magnetic unit-cell and thus cannot be explained by the present four-sublattice model.
The field dependence and the selection rules of $\eta_{11}$, the only mode found experimentally in the four-sublattice commensurate phase IV, are described by the  model, Fig.\,\ref{Fig:Magnetic_spectra}(c) and \ref{Fig:Magnetic_spectra}(f).

There are two resonances in the vicinity of the $\nu_1$ mode  as indicated by blue symbols in Fig.\,\ref{Fig:Magnetic_spectra}(b) and (e).
These two modes have a well-defined selection rule, \bwu{x}.
Because they are at low frequency but not described by the mean field model we assign them to impurity modes.

The exchange parameters obtained by fitting the mean-field model to THz spectra are presented in Table\,\ref{tabel:exchange parameters}. 
Our model also reproduces the magnetization for commensurate phases I and IV, Fig.\,\ref{Fig:Magnetization}.
The canting angle of spins  given by the parameters of the current work, Table\,\ref{tabel:exchange parameters} and Eq.\,(\ref{equation:spins_angle}), is $\pm\theta=8.1^\circ$ in zero field, in good agreement with the value determined by elastic neutron scattering, $(7.8\pm2.6)^\circ$, as reported in Ref.\,[\onlinecite{Toft-Petersen2011,Jensen2009_12}].


\section{Discussion}
\subsection{One-magnon excitations}

The four sub-lattice mean-field model describes four magnons $\nu_1$, $\nu_2$, $\nu_3$, and $\nu_5$, among which $\nu_1$ and $\nu_2$ can be identified as $\Gamma$-point magnon modes observed in the INS spectra \cite{Jensen2009}, whereas the $\nu_5$ resonance has also been detected by the Raman spectroscopy \cite{Fomin2002}.

The zero-field frequencies of $\nu_{1}$ and $\nu_{2}$ are related to the single-ion anisotropies $\Lambda_x$ and $\Lambda_y$, respectively. 
Furthermore, the selection rules for the $\nu_{1}$ and $\nu_{2}$  suggest that they are anisotropy-gapped magnons, since in both cases the magnetic dipole moment oscillates perpendicular to the corresponding anisotropy axis, along $y$ for the $\nu_{1}$ mode and along $x$ for $\nu_{2}$ in zero field. 
Moreover, the mean-field model  reproduces the rotation of the magnetic dipole moment of  $\nu_{1}$ ($\nu_{2}$)  towards the $z$ axis in increasing magnetic field  \bparal{y} (\bparal{x}).
The reappearance of $\nu_{1}$ in phase IV, marked as $\eta_{11}$, is also predicted by the model.

The frequencies of $\nu_{3}$ and $\nu_5$ depend strongly on the weak $J_{xy}$ and $J_{xz}$ exchange interactions connecting the two AF systems, $\{\mathbf{S}_{1},\mathbf{S}_{4}\}$ and $\{\mathbf{S}_{2},\mathbf{S}_{3}\}$. 
While the FM $J_{xz}$ only shifts the average frequency of $\nu_{3}$ and $\nu_{5}$, the AF $J_{xy}$ affects the difference frequency. 
The zero-field selection rules of these excitations -- magnetic dipole moment along $z$ for $\nu_{3}$ and the absence of magnetic-dipole activity of $\nu_{5}$ -- are reproduced by the model.

Our model does not describe the incommensurate phases II, III and V. 
However, it reproduces the frequency of the lowest $\eta_{11}$ mode in the commensurate phase IV, Fig.\,\ref{Fig:Magnetic_spectra}(c).

\begin{figure}
	\includegraphics[width=0.47\textwidth]{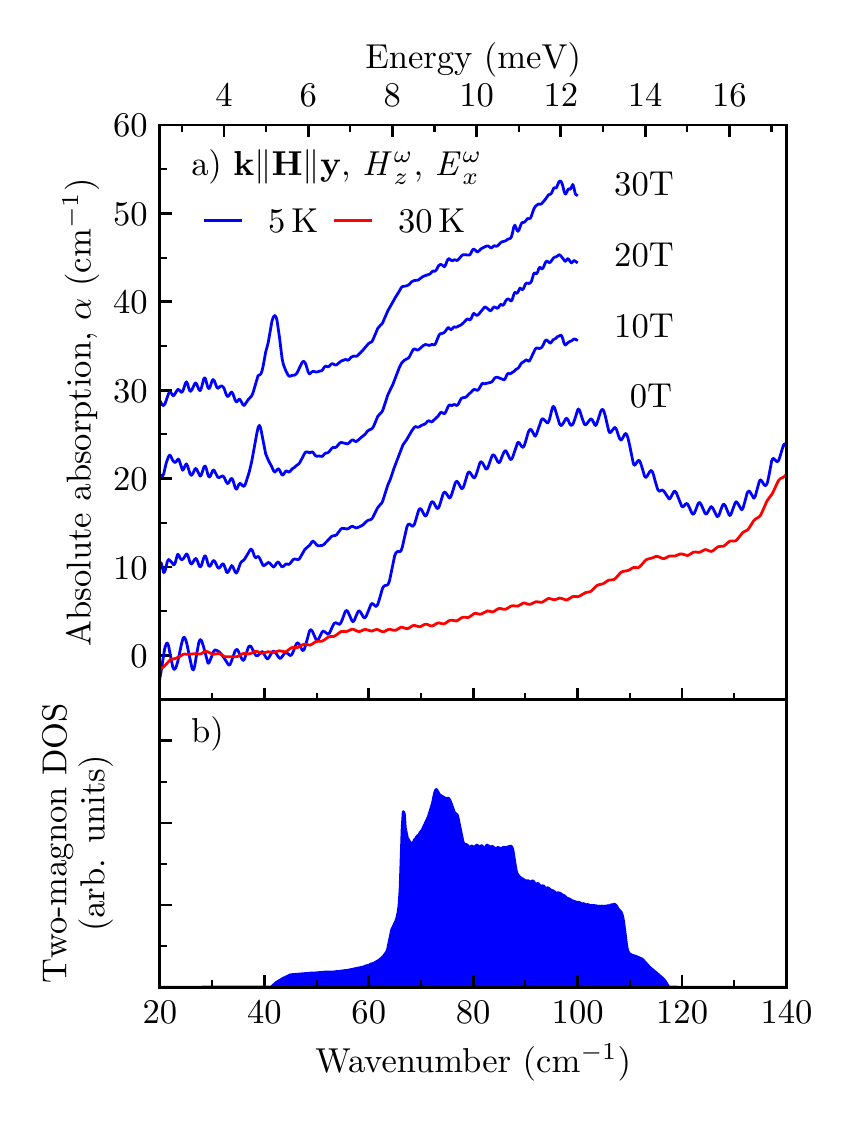}
	\caption{\label{Fig:Double_magnon}(a) Two-magnon excitation band $\nu_7$ in \lnpo/
		as observed in the experiment (blue spectra). (b) Calculated two-magnon density of states. 
		The two-magnon absorption is absent in the paramagnetic state, red line in panel (a), measured in 0\,T at 30\,K. 
		All  spectra are first shifted to zero absorption at 20\wn/ and then  a constant shift proportional to the field is added.
		The spectra in magnetic field were measured   up to 100\wn/.
	}
\end{figure}

\subsection{Two-magnon excitations \label{sec:TwoMagnon}}

Two-magnon excitations appear in the absorption spectra when one absorbed photon creates two magnons with the total $\mathbf{k}$ vector equal to zero \cite{Richards1967}.
The two-magnon absorption is  the  strongest where the density of magnon states is the highest, usually at the Brillouin zone boundary.
Since the product of the two spin operators has the same time-reversal parity  as  the electric dipole moment, the two-magnon excitation by the electric field is allowed; this mechanism usually dominates over the magnetic-field induced two-magnon excitation \cite{Richards1967}.

The broad absorption band between 60 and  115\wn/, shown in Fig.\,\ref{Fig:Double_magnon}(a), appears below \TN/ and is \ewu{x} active.
A similar excitation band  observed by Raman scattering was attributed to  spin excitations \cite{Fomin2002}.
In  another olivine-type crystal LiMnPO$_4$, a  broad  band in the Raman spectrum was assigned to a two-magnon excitation with the line shape reproduced using the magnon density of states (DOS) \cite{Filho2015}.

We calculated the magnon DOS  numerically on a finite-size sample of $4\times4\times4$ unit cells with 256 spins using the model represented by  Eq.\,(\ref{equation:model}) but extended by the $J_y$ and $J_z$ couplings shown in Fig.\,\ref{Fig:Magnetic_cell}. 
The two-magnon DOS was obtained by doubling the energy scale of the single-magnon DOS and is shown in Fig.\,\ref{Fig:Double_magnon}(b).
Since the observed broad absorption band emerges in the energy range of the high magnon DOS we assign this absorption band to a two-magnon excitation. 

There is another dominantly  electric-dipole active spin resonance  $\nu_6$ at 66.5\wn/ not reproduced by our mean-field model.
Since the frequency of the electric-dipole active $\nu_6$ mode is at the maximum of the two-magnon DOS and in a magnetic field, $\vect{H}\parallel \vect{z}$,  splits into a lower and an upper resonances with effective $g$ factors $g_-=4.24\pm0.07$ and  $g_+=4.00\pm 0.04$, i.e two times larger than that expected for one spin-flip excitation, we interpret the $\nu_6$ resonance as a two-magnon excitation. 
The singular behavior in the DOS coinciding with the $\nu_6$ resonance peak corresponds to the flat magnon dispersion along the R--T line in the Brillouin zone. 
This mode is weakly magnetic-dipole active  as well and is therefore  a ME resonance. 

The magnetic field dependence of the two-magnon excitation $\nu_{6}$ was modeled by calculating the field dependence of the magnon DOS.
The result is shown in Fig.\,\ref{Fig:Magnetic_spectra}.
The splitting of the resonance in magnetic field is observed only for \bparal{z} and is   reproduced by the model calculation. 
In the calculation $J_y$ was set to 0.65 meV to reproduce the instability of phase I at 12 T, while $J_z$ =0.16 meV was used to reproduce the zero-field frequency of $\nu_{6}$.
The magnitude  of $J_y$ and $J_z$ is similar to  the ones in  INS studies \cite{Toft-Petersen2011,Jensen2009,Li2009Ni} but $J_z$ has the opposite sign. 
In high-symmetry cases the electric-dipole selection rules of two-magnon excitations can be reproduced by group theoretical analysis \cite{Loudon1968}, but the low magnetic symmetry of \lnpo/ hinders such an analysis. 
Nevertheless, it is expected that the $\Delta S=0$ two-magnon continuum $\nu_7$ has different selection rules than the $\Delta S=2$ two-magnon excitation  $\nu_6$ due to their different symmetry.

\section{Conclusions}

We measured the magnetic field dependence of THz absorption spectra in various magnetically ordered phases of \lnpo/.
We have revealed a variety of spin resonance modes: three  magnons, an electromagnon, an electric-dipole active two-magnon excitation band, and a magnetoelectric two-magnon excitation. 
The abrupt changes in the magnon  absorption spectra coincide with the magnetic phase boundaries in \lnpo/.  
The magnetic dipole selection rules for magnon absorption and the magnetic field dependence of magnon frequencies in the commensurate magnetic phases are described with  a mean-field spin model. 
With this model the additional information obtained from the magnetic field dependence of mode frequencies allowed us to refine the values of exchange couplings, single-ion anisotropies, and \DM/ interaction. 
The significant differences found in magnetic interaction parameters compared to former studies are the opposite sign of $J_z$ exchange coupling, the smaller values of the $J_{xy}$ exchange coupling and the $\Lambda_x$ and $\Lambda_y$ anisotropies.
The mean-field model did not explain the observed magneto-electric excitation $\nu_4$ and the spin excitations in the incommensurate phases. 
In the future, more about the magneto-electric nature of \lnpo/ spin excitations can be learned from non-reciprocal directional dichroism studies as in \lcpo/ \cite{Kocsis2018}.

\section{Acknowledgments}
The authors are indebted to L\'aszl\'o Mih\'aly and Karlo Penc for valuable discussions.
This project was supported by institutional research funding IUT23-3 of the Estonian Ministry of Education and Research, by the European Regional Development Fund project TK134, by the bilateral program of the Estonian and Hungarian Academies of Sciences under the Contract No. NKM-47/2018, by the Hungarian NKFIH Grant No. ANN 122879, by the BME Nanotechnology and Materials Science FIKP grant of EMMI (BME FIKP-NAT), and by the Deutsche Forschungsgemeinschaft (DFG) via the Transregional Research 
Collaboration TRR 80: From Electronic Correlations to Functionality (Augsburg-Munich-Stuttgart).
D. Sz. acknowledges the FWF Austrian Science Fund I 2816-N27 and V. K. was supported by the RIKEN Incentive Research Project.
High magnetic field experiments in HFML were funded by EuroMagNET under the EU Contract No. 228043.

\bibliographystyle{apsrev_no_DOI_ISSN}

\end{document}